\documentclass[fleqn,10pt]{wlscirep}
\usepackage[utf8]{inputenc}
\usepackage[T1]{fontenc}

\title{Twisting harmonics: Transfer of orbital angular momentum in solid-state high-harmonic generation}
\usepackage{xcolor}

\author[1]{Debobrata Rajak}
\author[2]{Bikash Kumar Das}
\author[1]{Rajaram Shrestha}
\author[1]{Bálint Kiss}
\author[1,3]{Eric Cormier}
\author[4]{Carmelo Rosales-Guzman}
\author[2,5,6]{Stephan Fritzsche}
\author[7,8,9]{Qiwen Zhan}
\author[10]{Wenlong Gao}
\author[10,*]{Camilo Granados}

\affil[1]{ELI-ALPS, ELI-Hu Non-Profit Ltd., Szeged, Hungary.}
\affil[2]{Theoretisch-Physikalisches Institut, Friedrich-Schiller-Universität Jena, Max-Wien-Platz 1, D-07743 Jena, Germany}
\affil[3]{Laboratoire Photonique Numérique et Nanosciences (LP2N), UMR 5298, CNRS-IOGS-Université Bordeaux, 33400 Talence, France}
\affil[4]{Centro de Investigaciones en Óptica, A.C., Loma del Bosque 115, Colonia Lomas del campestre, 371507 León, Gto., Mexico}
\affil[5]{Helmholtz-Institut Jena, D-07743 Jena, Germany}
\affil[6]{GSI Helmholtzzentrum f\"ur  Schwerionenforschung GmbH, D-64291 Darmstadt, Germany}
\affil[7]{School of Optical-Electrical and Computer Engineering, University of Shanghai for Science and Technology, Shanghai 200093, China}
\affil[8]{Zhejiang Key Laboratory of 3D Micro/Nano Fabrication and Characterization, Department of Electronic and Information Engineering, School of Engineering, Westlake University, Hangzhou 310030, China}
\affil[9]{International Institute for Sustainability with Knotted Chiral Meta Matter (WPI-SKCM2), Hiroshima University, Higashihiroshima, Hiroshima 739-8526, Japan}
\affil[10]{Eastern Institute of Technology, Ningbo 315200, China}

\affil[*]{cagrabu@eitech.edu.cn}

\keywords{Attosecond sciences, Structured light, High-order harmonic generation}

\begin{abstract}
Although solid-state platforms underpin modern electronics, little is known about how intense ultrashort light pulses carrying orbital angular momentum (OAM) interact with solids. This gap persists even though, for more conventional light–matter interactions, the complex underlying electron dynamics can often be confined to a single Brillouin zone and described well within the dipole approximation. Previous studies were restricted to nonlinear, perturbative regimes, largely because the generation of intense ultrashort vortex pulses, particularly in the mid-infrared spectral regime, has remained a long-standing challenge. Consequently, the role of structured light in driving nonlinear, non-perturbative processes in solids, and the associated transfer of angular momentum during these interactions, has not been systematically explored. Here, we investigate solid-state high-harmonic generation (HHG) driven by intense ultrashort structured light using a versatile experimental approach applicable to different materials and geometries. We demonstrate that the OAM of the driving field is coherently transferred to the emitted harmonics. In particular, we show that the OAM is conserved independently of the crystal symmetry, the range of electronic interactions, and the presence of strong spin–orbit coupling. These results establish OAM-resolved HHG as a robust framework for characterizing and controlling angular momentum transfer in solid-state HHG and open new avenues for structured-light-driven quantum technologies and topological materials investigations.
\end{abstract}

\begin{document}

\flushbottom
\maketitle
% * <john.hammersley@gmail.com> 2015-02-09T12:07:31.197Z:
%
%  Click the title above to edit the author information and abstract
%
\thispagestyle{empty}

% \noindent Please note: Abbreviations should be introduced at the first mention in the main text – no abbreviations lists. Suggested structure of main text (not enforced) is provided below.

\section*{Introduction}

Conservation laws, arising from the symmetries of physical systems, form the foundation for predicting and understanding physical phenomena\cite{Nother}. A familiar example is the conservation of energy in a closed system, which follows from the time-translation symmetry of the system. In high-harmonic generation (HHG)- a coherent, nonlinear, non-perturbative process- energy conservation and the induced dipole dynamics typically lead to emission of radiation at frequencies that are integer multiples of the fundamental frequency, i.e. $\omega_q = q\omega_0$, where $\omega_0$ is the driving field's central frequency and $q$ is the harmonic order\cite{Lewenstein}. This frequency up-scaling law has been observed for HHG in atomic gases, solids, and liquids \cite{HHG_solids, HHG_solids2,HHG_gases, liquids}. Symmetry breaking in the driving laser–matter system, for instance using two-color laser fields, allows the appearance of even harmonics alongside odd harmonics\cite{Breaking_spatial}. Similarly, the rotational invariance of an electromagnetic field dictates the conservation of orbital angular momentum (OAM) of light in gas-phase HHG \cite{Allen_OAM,HHG_OAM_EXP}. Experiments using vortex beams- also known as phase singular beams, spatially structured beams or helical beams- show that the generated harmonics inherit the structural driving field's characteristics due to the coherent nature of the HHG process \cite{HHG_OAM_EXP}. Vortex beams typically possess a doughnut-shaped intensity distribution and a helical phase front, defining an OAM of $l_{0}\hbar$ per photon. When such beams drive the HHG process in gases, the harmonic OAM, $l_q$, scales linearly with both the harmonic order, $q$, and the fundamental beam's OAM, $l_0$, i.e., $l_q=ql_0$, making it a conserved quantity \cite{HHG_OAM_EXP}. This scenario is well described within the dipole approximation, owing to a large disparity of spatial scales involved in the interaction i.e., between the electron excursion distance and the transverse dimensions of the vortex beam or its wavelength \cite{Corkun1, Bikash_POV}. 

Combining structured light beams with nonlinear, non-perturbative optical processes has significantly advanced our ability to explore and control matter\cite{Corkun1,HHG_OAM_EXP,HHG_OAM_EXP2,HHG_OAM_EXP3,HHG_OAM_EXP4,HHG_OAM_EXP5,Stephan1}. This progress has been made evident in the trapping of particles\cite{Theory_Trapping,Exp_Trapping} and the generation of twisted attosecond pulses\cite{AttoOAM}. Another interesting application that demonstrates our control over light is the so-called self-torque of light \cite{SelfTorque}, where the OAM no longer remains static, rather it changes dynamically in the sub-fs time scale. In solids, however, several physical mechanisms, where the electron phase and the crystal symmetry are fundamental, can challenge the ideal transfer of optical OAM. Crystal anisotropy, interfaces, and crystal defects can create pathways for OAM redistribution and mode mixing via angular momentum exchange with the lattice \cite{OAM_Constrained}. In practice, however, technical limitations in generating ultrashort structured beams in the mid-infrared (MIR) spectral regime have constrained experimental studies largely to the perturbative regime \cite{SHG, Nonlinear_SL, Flat_OAM}. Early attempts to extend studies beyond low-order harmonics relied on the interferometric detection of harmonics in ZnO\cite{OAM_solids}, but were fundamentally limited by phase sensitivity, detection efficiency, and restricted harmonic order. Furthermore, the interaction of circularly-polarized light with a thick gallium selenide crystal was exploited to generate vector structured harmonics via the spin–orbit coupling and to demonstrate that the crystal symmetry is linked to the emission of structured harmonics\cite{SAM_OAM_GaSe}. Thus, extending those measurements to different solid-state materials and high harmonics is fundamental to elucidate the effect of solid-state material's intrinsic properties on OAM conservation. 

Consequently, understanding HHG in solids driven by vortex beams is a key open question. Recent studies have shown that vortex beams can induce helicoidal dichroism in several systems \cite{Helical_dichro, Helical_dichro2, Helical_dichro3}, while interactions beyond the dipole approximation have been observed in atoms and molecules \cite{Beyond_diple,AlbarEsra}. In Ref.~\citenum{AlbarEsra}, for instance, it was demonstrated that the inclusion of the quadrupole moment in the vortex-matter interaction breaks the symmetry of the system and induces even harmonics in the HHG spectrum. Vortex beams provide a versatile tool to distinguish material responses to opposite OAM helicities, analogous to circular dichroism associated with spin angular momentum\cite{Rosales-Guzman2012, KaynForbes2019, Brullot2016}. In chiral or symmetry-broken solids, OAM-dependent absorption (demonstrated in gases \cite{TransferOAM}) and excitation pathways could enable ultrafast OAM-selective photo-detection, all-optical switching, and high-density information processing \cite{Forbes, OAM_Information, RMP_VorCondence}. Furthermore, the use of topological light with a large number of photons to control light-matter interaction and investigate topological materials was recently proposed\cite{ExtremeOAM}, opening future routes for the application of helical beams. These advances highlight the potential of OAM as a probe of symmetry-dependent dynamics, while underscoring the need to rigorously test OAM conservation in the strong-field solid interaction regime.

Here, we explore systematically how vortex beams interact with solids by experimentally demonstrating the conservation of OAM across a broad range of materials and HHG geometries. We first generate ultrashort, few cycles MIR structured light. We apply a robust but still simple static-optics approach (Fig.~\ref{ExpSet} top), in order to drive the HHG process in the transmission geometry and for the following materials in the bulk form: (1) a-cut Zinc Oxide (ZnO), a wide-bandgap semiconductor, (2) Gallium Selenide (GaSe), a layered Van der Waals  semiconductor, and (3) Magnesium Oxide (MgO), a wide-bandgap semiconductor. We further investigate how OAM is transfer from the driving field to non-perturbative harmonics in the reflection geometry for silicon (Si), an amorphous material and monolayer Tungsten Diselenide (WSe$_2$), a transition metal dichalcogenide. We demonstrate that the OAM conservation law is robust across all these widely differing crystalline symmetries, band gaps, and interaction mechanisms. In all our measurements, the OAM of the generated harmonics increase strictly linear with the harmonic order and the driving field's topological charge (TC), with no detectable deviations. To interpret these results, we develop a semi-empirical model within the dipole approximation that accurately captures the observed experimental features. Although our results establish the robustness of OAM conservation during strong-field laser–solid interactions, further measurements with other target material, such as metals and topological insulators, will be crucial to fully map the boundaries of this fundamental principle.  

\section*{Results}

\subsection*{Experimental measurements}

\begin{figure}
\includegraphics[width=1\linewidth]{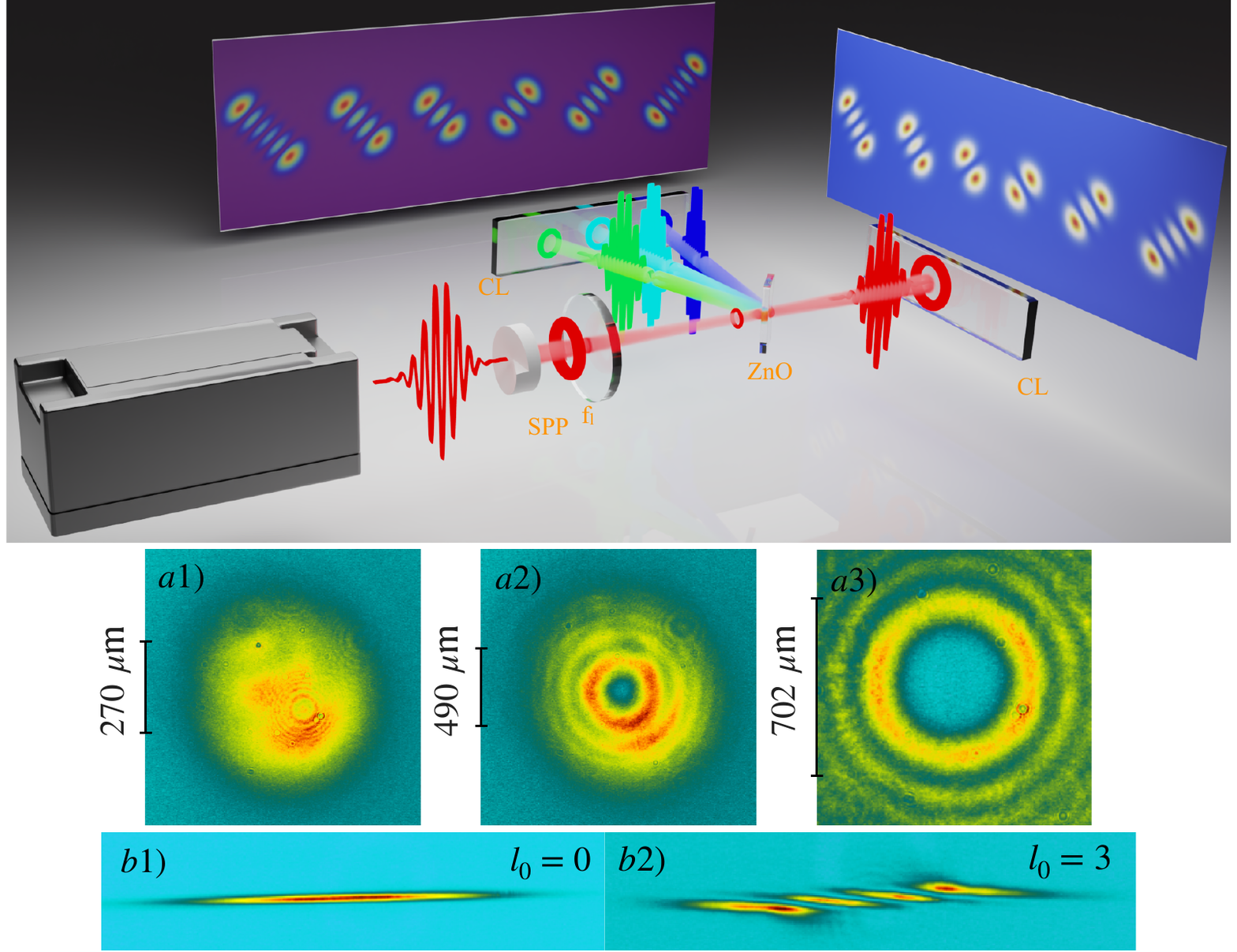}
\caption{ (Top) Simplified experimental setup. A fundamental Gaussian pulse first passes through a spiral phase plate (SPP)  and creates a LGV beam which is focused upon different target crystals, for instance ZnO, to drive the HHG process. The harmonic vortex beams are produced either in transmission or reflection geometry, depending on the crystal angle relative to the incoming LGV beam. Here, we exemplify the reflection geometry and represent different harmonic vortices by different colors, which are separated by band-pass filters during the experimental measurements. Furthermore, the spatial characterization of both the fundamental and harmonic vortices as well as HG profiles is performed via imaging cameras. The TC of the beams is measured by performing a mode transformation with a cylindrical lens (CL), which is positioned between the target crystal and the imaging camera. In (a1), we show the fundamental Gaussian beam's spatial intensity distribution, whereas in (a2) and (a3), we display the intensity distributions of LGV beams for two different TC values. Note that the SPP does not produce LGV beams with zero radial index. In (b1) and (b2), we compare the mode-transformed intensity distributions for the Gaussian beam and LGV beam with a TC $l_0=3$.}
\label{ExpSet}
\end{figure}

To validate the OAM conservation law in solid-state HHG, two fundamental questions must be addressed: First, How does the OAM scale with with harmonic order? Second, to what extent is the resultant harmonic OAM sensitive to the material's intrinsic properties? To answer these questions, we first generated linearly polarized femtosecond (fs), mid-infrared (MIR) Laguerre–Gaussian vortex (LGV) beams, and then focused these beams onto different target crystals to induce nonlinear effects. A simplified representation of the experimental setup at the Extreme Light Infrastructure - Attosecond Laser Pulse Source (ELI-ALPS, MIR laser source) facility is shown in Fig.~\ref{ExpSet} (top), which outlines the experimental details required to generate harmonic vortices (see also the Methods section). The initially fs Gaussian beam was transformed into an LGV beam using a spiral phase plate (SPP, Vortex Photonics, Munich). The transverse intensity distribution and the TC were characterized for both, fundamental Gaussian and LGV beams, using a MIR camera (IRC912 from IRCameras). Additionally, we measured the vortex beam size at the focus of the off-axis parabolic mirror, its temporal duration and the maximum energy per pulse, which resulted in $w\approx 65~\mu$m, $\tau_{p}\approx 47$~fs, and $E_{p}\approx 20$~$\mu$J, respectively. This energy corresponds to a peak intensity per pulse $I_{\text{peak}}=6.4\times 10^{11}$ W/cm$^2$ calculated at the focus position. The temporal duration of the vortex pulse was measured via the frequency-resolved optical gating (FROG) technique \cite{frog}. The repetition rate of the laser system is 100~kHz.   

We characterized the fundamental Gaussian and LGV fields spatially and analyzed their TC by measuring the near-field intensity distributions. In Figs.~\ref{ExpSet}~(a1)-(a3), we present the transverse intensity distributions of the fundamental Gaussian ($l_{0}=0$), and LGV beams with TCs $l_0=1$ and 3, respectively. To measure the TC of the beams, we used a cylindrical lens (CL). The measured intensity distributions are shown in Figs.~\ref{ExpSet}~(b1) and (b2), for the fundamental beams with $l_0=0$ and 3, respectively. The CL transforms the input vortex mode into a tilted Hermite-Gaussian (HG) distribution based on the astigmatic mode conversion principle~\cite{Beijersbergen1993,Padgett2002}, which is essentially an optical Fourier transformation (OFT) operation. From the tilted HG intensity distribution, it is possible to extract the vortex mode's TC by counting the number of intensity minima, $n$, between the bright lobes \cite{Beijersbergen1993,Padgett2002}. For the fundamental Gaussian beam, the mode conversion produces an elongated Gaussian-like profile with no minima, $n=0$, and confirms a TC of zero ($l_0=n$). For the LGV beam, in contrast, the mode conversion generates the expected tilted HG structures, as demonstrated in Fig.~\ref{ExpSet}~(b2). In the resulting HG profile for the LGV beam, we observe three minima, $n=3$, corresponding to $l_0=3$, which is consistent with the SPP used to imprint the TC. Similarly, for the LGV mode with $l_{0}=1$, we observe one intensity minimum between the two bright lobes (see the Supplementary Material).  

\subsection*{Transmission geometry}

\begin{figure}[ht!]
\includegraphics[width=1\linewidth]{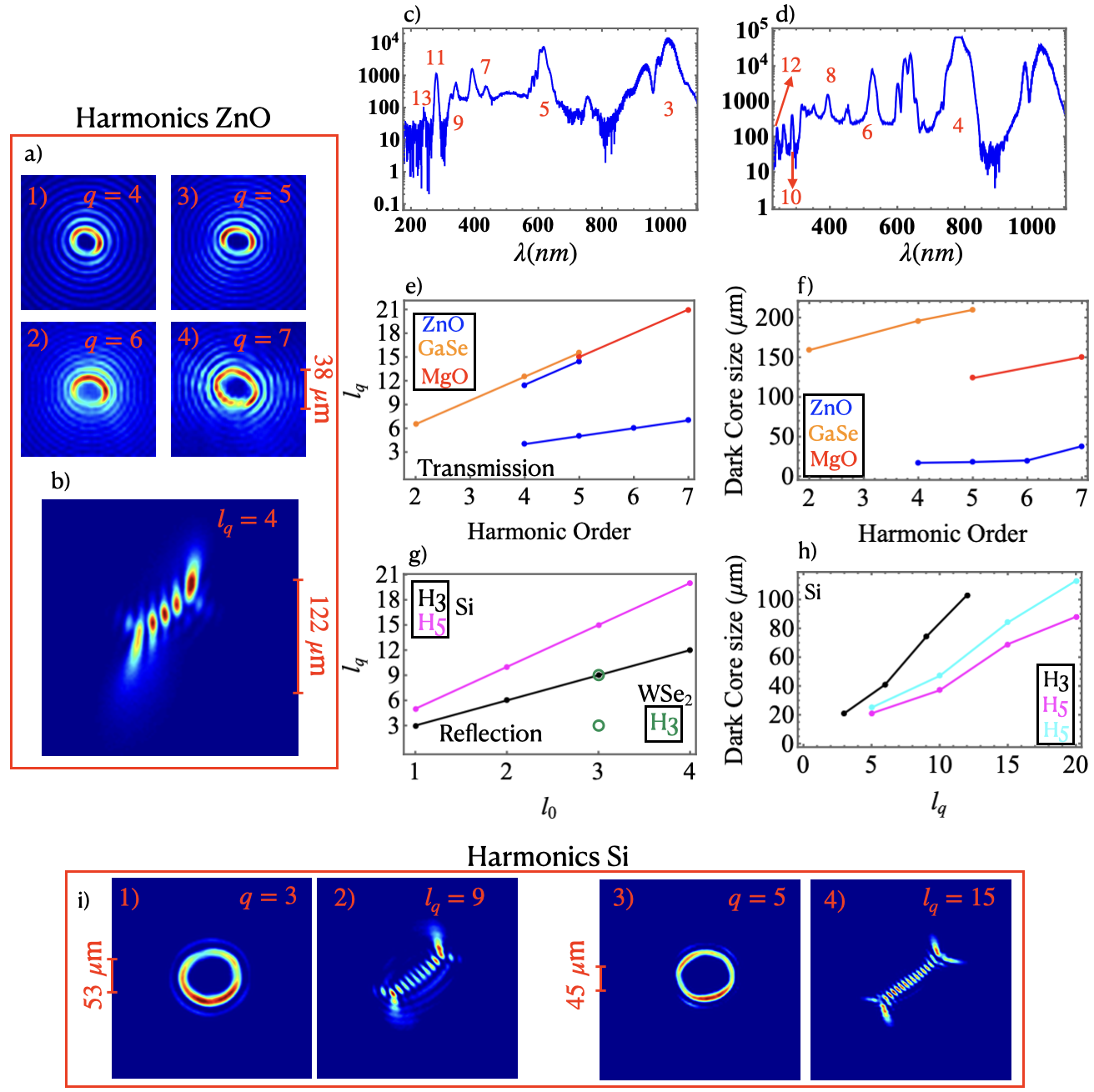}
\caption{Harmonic vortices, OAM up-scaling law, and beam sizes. In (a), we show the measured intensity distributions of the harmonic vortices $q=3,4,5$ and 7 generated in the ZnO crystal with a fundamental LGV beam carrying a TC $l_{0}=1$. In (b), the resulting HG intensity distribution for the $q=4$ harmonic vortex beam, demonstrates that the TC follows: $l_q=ql_0$. In (c) and (d), the spectral characterization of the generated harmonic vortices (both, even and odd orders) in ZnO. In (e), the linear up-scaling law describing the harmonic's TC, $l_q$, as a function of the harmonic order, $q$, for GaSe, ZnO and MgO. In (f), harmonic vortices dark-core size as a function of the harmonic order, demonstrating its increment for increasing $q$ values, for the same solids. In (g), the OAM up-scaling law for silicon and WSe$_2$ (green open circles), is shown for validity for the reflection geometry. In (h), the enhancement in the size of the dark-core as a function of the harmonic's TC, $l_q$. In (i), harmonic vortices and HG intensity distributions for harmonics $q=3$ and 5 generated in silicon. For both harmonic vortices, the fundamental beam carries a TC of $l_0=3$, resulting in $l_q=9$ and 15, respectively. Note that the error assigned to all the dark-core values corresponds to 12~$\mu$m, which is the pixel pitch in the detection MIRC.}
\label{HHGV}
\end{figure}

We begin our investigation by generating harmonic vortices in a transmission geometry. In this geometry, the LG vortex beam passes through the target crystal, with harmonics generated at the rear surface of the crystal to avoid re-absorption. This geometry for HHG is highly efficient and allows the process to occur even at normal incidence. More importantly, this geometry allows us to determine the impact of propagation effects on the OAM of the emitted harmonics. A central prediction for OAM in the HHG process is that the TC of harmonic vortices must scale with the harmonic order following the up-scaling law $l_q=ql_0$, in a clear analogy with the frequency up-scaling law $ \omega_q=q \omega_0$. We verified the OAM up-scaling by generating several harmonic vortices in a 200~$\mu$m-thick ZnO crystal driven by an LG vortex beam with a TC of $l_0=1$ and in ambient air. The ZnO crystal possesses a direct bandgap of $E_{g}=3.37$ eV at room temperature. In Figs.~\ref{HHGV} (a1)-(a4), we display the measured transverse intensity distributions of the resulting harmonic vortices ($q=4$ to 7). We obtained identical results with a 90~$\mu$m-thick ZnO crystal too, confirming the robustness of the effect.

To extract the TC of the harmonic vortices, we counted the number of minima from their corresponding HG intensity distribution, which resulted in $n=4,5,6$ and 7 (see the Supplementary Material). Since the number of minima, $n$, is exactly equal to the harmonic order $q$, we can conclude that the TC of harmonic vortices scales linearly with the harmonic order, i.e., $l_q=q$, since $l_{0}=1$ in this case. In Fig.~\ref{HHGV} (b), we show an example of the resulting HG intensity distribution, for the fourth harmonic vortex. It is clear from the figure that the number of minima is $n=4$, resulting in a TC of $l_q=4$. Furthermore, in Figs.~\ref{HHGV} (a1)–(a4), we observe harmonic vortices with multiple concentric off-axis rings surrounding the central dark-core. This occurs because the fundamental LG vortex beam, shown in Fig.~\ref{ExpSet} (a2), also contains multiple off-axis rings, though less than what we see in the harmonic vortices. This reflects the fact that the HHG process in solids amplifies mode imperfections of the fundamental OAM beam, thereby, reducing the mode purity of the generated harmonics. Figures~\ref{HHGV} (a1)–(a4) show the size of dark-core and the radius of the maximum intensity of harmonic vortices increasing with the harmonic order, although none of these changes are significant. This increased beam size of the harmonics reflects that they all are measured at a fixed focal plane. 

Furthermore, in Figs.~\ref{HHGV} (c) and (d), we show the measured HHG spectra for ZnO that span the harmonic orders from $q=3$ to $q=13$. In particular, the spectrum presented in Fig.~\ref{HHGV} (d) includes both even and odd order harmonics. Due to limitations in the detection system, our OAM measurements do not include harmonic orders above eight. We note that in an a-cut (plane $\langle 11\bar{2}0 \rangle$) ZnO crystal, based on how the input laser polarization direction is oriented with respect to the crystal axis, the spatial symmetry can either be preserved or broken. If the spatial symmetry is preserved, only odd harmonics appear in the harmonic spectrum. However, if the spatial symmetry is broken, odd an even harmonic order will occur both in the (harmonic) spectrum. On the other hand, in a c-cut ZnO crystal, the symmetry is always preserved, irrespective of how one orients the laser polarization direction with respect to the crystal axis. Most importantly, we observe that the same up-scaling law for OAM holds irrespectively of the spatial crystal symmetry. Before entering the camera, the fundamental and harmonic vortex beams were separated using a borosilicate crown glass (BK7) plate. Altogether, these results demonstrate the OAM conservation in solid-state HHG in ZnO, in agreement with the earlier work \cite{OAM_solids}. Our implemented experimental technique, however, is much simple and versatile since it implements only static optical elements to measured the different structured light properties (see the Supplementary Material for a complete set of HG intensity distribution measurements).

To confirm the linear OAM up-scaling law in solid-state HHG, we repeat the measurements using a higher TC of the driving-field, $l_0=3$, and various other materials: GaSe, a non-centrosymmetric crystal characterized by a bandgap of $2.1$ eV, and MgO, characterized by a direct bandgap of $7.8$ eV. From the TC measurements we confirm that the OAM up-scaling law is also preserved. A summary of all measured TCs as a function of the harmonic order and for the three different solid materials, is shown in Fig.~\ref{HHGV} (e). To clearly show the scaling behaviour of the harmonic vortices TC, we slightly offset the data points for the MgO and ZnO crystals horizontally since they coincide with the data points for the GaSe crystal. These results clearly depict the linear up-scaling law for the OAM i.e., $l_q=ql_0$. It is important to highlight that although ZnO, GaSe, and MgO crystals possess fundamentally different crystal structures, symmetries, and minimum bandgap, the OAM conservation law appears to be exactly the same for all of them. 

The nonlinear response of different materials to the helical beam is also seen form the increased size of the central central dark-core with the harmonic order. Figure~\ref{HHGV}~(f) displays the dark-core size as a function of the harmonic order for the different materials. It is clear from the figure that the size of the dark-core expands monotonically with increasing harmonic orders and that the behaviour is different for different materials. This trend was reported earlier for the ZnO crystal \cite{OAM_solids}. A comparison between the dark-core size of the harmonic orders $q=4$ and $q=5$ generated from GaSe and ZnO crystals reveals that the growth of the dark-core size is faster in the case of GaSe. Similar conclusions can be drawn when comparing the dark-core size of harmonic orders $q=5$ and $q=7$ generated from MgO and ZnO crystals. The distinct functional dependencies observed in our experiments across materials reflect their different nonlinear responses to the structured light. These features were explained earlier, for ZnO, using a theoretical framework introduced in Ref.~\citenum{Granados_2025}, which combines the semiconductor Bloch equations (SBEs) with the thin-slab model (TSM) to capture both the microscopic and macroscopic dynamics of LGV beams-driven HHG (see the Methods section). The theoretical work confirmed the role of the nonlinearity of the HHG process as the source of this particular effect. 

The measured HG intensity distributions demonstrate that OAM conservation holds for all the solid materials studied in the transmission geometry. Additionally, the HHG process in solid materials, we demonstrated that it is possible to produce harmonic vortex beams in the infrared-to-visible spectral ranges ($q=3, 4, 5, 6$ and 7 correspond to wavelengths $\lambda_{3}=1067$~nm, $\lambda_{4}=800$~nm, $\lambda_{5}=640$~nm, $\lambda_{6}=533$~nm, and $\lambda_{7}=457$~nm, respectively) with high TC values (e.g., $l_q = 21$ in the case of MgO and $l_q = 15$ for the GaSe case). Most importantly, because the fundamental beam's wavelength lies in the MIR spectra regime, we demonstrate that HHG driven by helical beams and the OAM transfer from the fundamental beam to the harmonic field can be studied in ambient air. The generation of harmonic vortices with large values of OAM in the IR-to-visible spectral regime, greatly enhances the potential applicability of ultrashort harmonic vortices in optical techniques aimed to achieve high spatial resolution, where vortex beams have been shown to offer significant advantages over fundamental Gaussian beams (TEM$_{00}$ modes)~\cite{STED_vortex}. Another key observation from the transmission geometry based HHG driven by LGV beams is that propagation effects fundamentally play no role in altering the OAM of the emitted harmonics. In other words, the OAM remains exactly the same in the near- (close to the crystal exit) and far-fields (at the detector).

\subsection*{Reflection Geometry}

To further test the OAM conservation law, we extended our experimental measurements to a reflection geometry. This geometry is crucial for understanding the influence of the material-dependent dipole-phase on the OAM conservation during the HHG process, as it dominates over contributions coming from the self-phase modulation (SPM), as reported earlier in Ref.~\citenum{Shambhu_Interferometry}. Moreover, the reflection geometry is highly relevant because it allows HHG from metals and other opaque materials and naturally provides access to surface-sensitive measurements. Verifying the OAM conservation in this geometry offers a stringent benchmark for the theoretical model (see Methods), since the surface emission allows the solid to be effectively approximated as a thin layer, minimizing propagation (specifically, the SPM contribution) and dephasing effects \cite{dephasing_effects,Shambhu_Interferometry, Granados_2025}. Together, the complementary material properties and geometries help establish a more complete picture of the OAM conservation in solid-state HHG and test the general applicability of the theoretical framework developed in Ref.~\citenum{Granados_2025}.

We performed the experiments on two different materials: (1) silicon (Si, 1-cm thick) an amorphous material with short-range ordering\cite{Gauthier_Polarization_Si}, characterized by an indirect bandgap of $1.6$ eV and a markedly different electronic structure compared to crystalline solids which typically display long-range ordering; and (2) monolayer WSe$_2$, characterized by a direct bandgap of $1.7$ eV and possesses one of the strongest spin–orbit couplings among transition-metal dichalcogenides (TMDs), measured at 513(10)~meV \cite{Dowben}. This property could, in principle, be coupled to the OAM of structured lights and affect the resulting harmonic vortices TC linear up-scaling. To generate harmonics in the reflection geometry, the samples were positioned at an angle of $\theta \approx$ 45$^{\circ}$ relative to the incident fundamental vortex beam. This angle was optimized to maximize the HHG throughput. The reflected fundamental and harmonic vortices were then directed to the detection system to characterize their transverse intensity distribution. 

The results in Fig.\ref{HHGV} (g) confirm the same linear OAM up-scaling law $l_q=ql_0$ for both Si and WSe$_2$. For a Si target, we observe a clear increase in the dark-core size with the harmonic's TC, as presented in Fig.\ref{HHGV} (h) for the third and fifth harmonic vortices. The apparent decrease in the dark-core size from the third to fifth harmonic (black and magenta lines in Fig.~\ref{HHGV}(h)) arise from low harmonic yields in the reflection geometry, which prevents measurements away from the focal plane. As a result, all harmonics were measured at their focal position, where higher-order harmonics (with shorter wavelengths) naturally produce smaller beam sizes, explaining the smaller dark-core sizes measured for the fifth harmonic. After correcting for this effect using a relative-size ratio extracted from measurements of $q=3$ and $q=5$ at the same position, the corrected dark-core size of the fifth harmonic shows an increasing trend relative to the third harmonic, as shown in Fig.~\ref{HHGV} (h) by the cyan line. 

Furthermore, an example of the transverse intensity profiles for the third and fifth harmonic vortices generated in Si is presented in Fig.~\ref{HHGV}~(i). These harmonics were generated with a fundamental LGV beam carrying a TC of $l_0=3$, resulting in harmonic vortices with TCs $l_q=9$ and 15, respectively. Their TC becomes evident in this figure by counting the minima, $n$, in the associated HG intensity distribution. A comparison between the harmonics reveals that the ring thickness decreases with increasing harmonic order. This harmonic vortex characteristic was also observed for ZnO crystal in Ref.~\citenum{HHG_solids}. Moreover, the distinct transverse intensity profile of the fifth harmonic generated from ZnO (in a transmission geometry) and Si (in a reflection geometry) reveals that the reflection setup reduces the number of off-axis rings present in the transverse intensity distribution, likely due to the low harmonic yield.

For a monolayer WSe$_2$ target, the efficiency of the HHG is low and restrict our measurements to the third harmonic only; nonetheless, it is possible to drive the HHG process with two different TC values of the fundamental beam, $l_0=1$ and 3, as shown in Fig.~\ref{HHGV} (g) with green open circles. In both cases, the resulting harmonic OAM is well-described by the linear up-scaling law $l_q=ql_0$. Thus, even strong spin-orbit coupling does not affect the OAM up-scaling law in this monolayer material. For WSe$_2$, we observe an enhancement in the beam size with increasing TC values of the fundamental beam, which is consistent with the trends observed in case of Si and GaSe (see the Supplementary Material).

Overall, the results obtained from the reflection geometry clearly demonstrate the OAM conservation at the surface of the materials studied here and show that harmonic vortices generated in the reflection and transmission HHG setup exhibit similar far-field spatial characteristics. This allows us to conclude that propagation and dipole phase effects do not play any fundamental role in the conservation of OAM in solids. However, changes in the transverse intensity distributions of different harmonics for different solid materials and HHG geometries reflect their different nonlinear response to the helical light. These findings firmly support the theoretical consideration of a solid-state material as an effective thin layer as proposed in the thin-slab model, and simplify the description of their nonlinear response to an input vortex beam\cite{Granados_2025}. Furthermore, the large spin-orbit interaction in WSe$_2$ seems not to modify its OAM response, as the conservation law remains intact. Extending such measurements to materials with topological surface states, such as Bi$_2$Se$_3$, and retrieving the phase of the emitted harmonic vortices will be crucial for providing greater insight into light–matter interactions \cite{RMP_VorCondence}. 

\subsection*{Theoretical model}

To characterize the nonlinear macroscopic response of a solid-state target to a vortex light field, we model their interaction by treating the solid target (crystals in our case) as a thin layer~\cite{Granados_2025} (see also the Methods section). This approximation is supported by two key experimental facts: (1) high-order harmonics are generated predominantly near the crystal's exit face, as emissions from within the bulk are effectively suppressed by harmonic re-absorption, which means for any harmonic generation experiment, we always work in these conditions, and (2) harmonic vortices produced in the reflection geometry exhibit similar far-field spatial features as those generated in the transmission geometry, as demonstrated for silicon. These two findings justify modeling the original target as an effective thin layer rather than as a bulk medium. Additionally, by positioning the thin layer exactly at the focal plane of the driving laser field, harmonic phase contributions arising from the focusing, i.e., the Gouy phase, and the curved wave front, i.e., the radial phase, of the driving beam can be disregarded, while the dipole phase is neglected because it only produces secondary effects in the electron dynamics and its contribution to the harmonic phase is at least an order of magnitude weaker than that of the helical phase contribution \cite{Granados_2025}. Also, we consider that the dipole approximation is valid in such an interaction, since the OAM is conserved, and the electron excursion distance (typically, a few nanometers depending on the driving field's peak amplitude and wavelength) is negligible as compared to the vortex beam's waist size ($\sim 65~\mu$m as measured in our experiment) and wavelength ($3200$~nm). This approximation effectively decouples the temporal and spatial responses. Consequently, the electron dynamics is governed by the temporal structure of the driving field, while the spatial profiles of the emitted harmonics follow the driving field's characteristic doughnut-shaped intensity distribution. 

\begin{figure}[h!]
\includegraphics[width=1\linewidth]{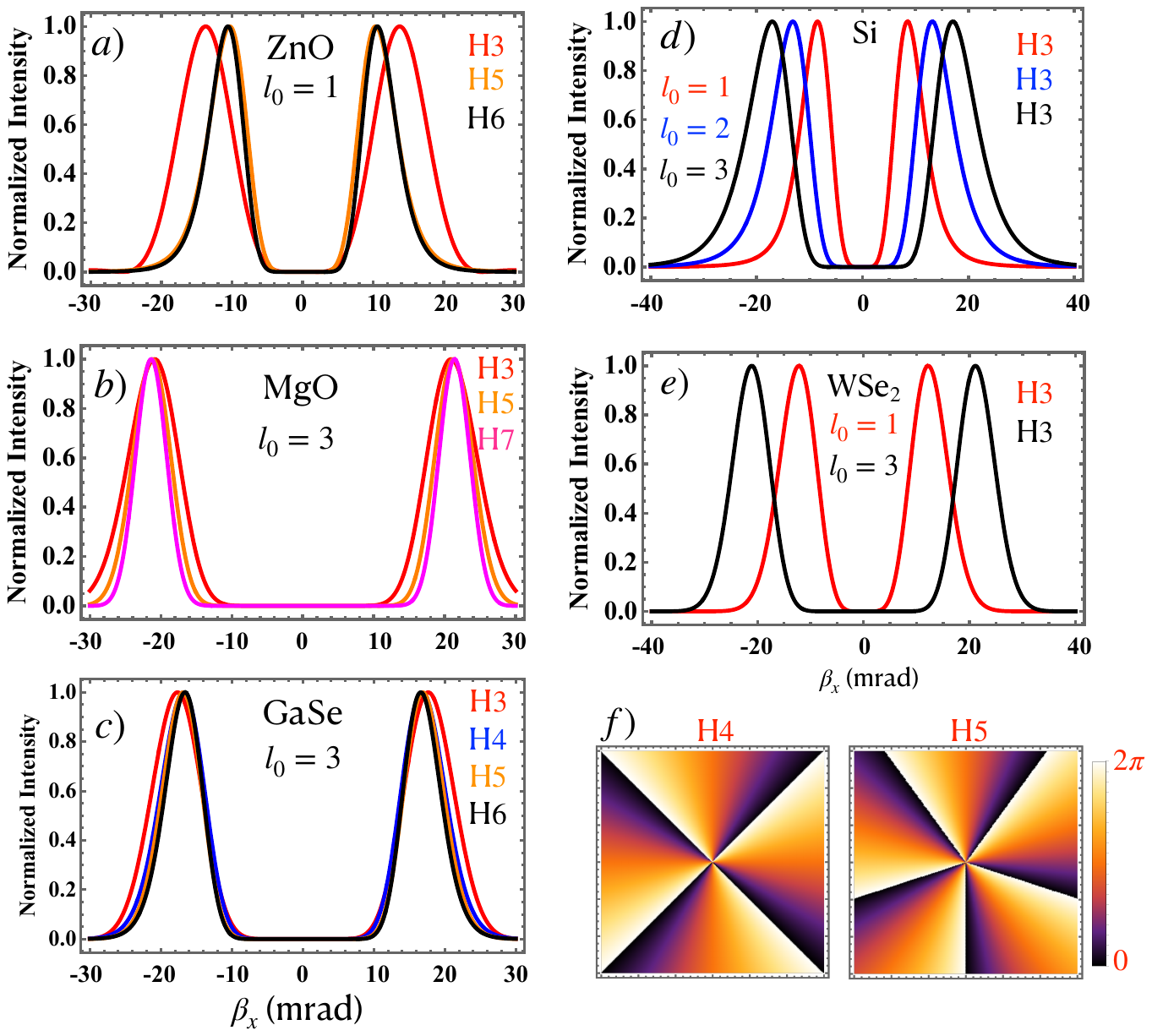}
\caption{Calculated far-field line intensity and phase profiles of the harmonic vortices for different materials and geometries. Results are shown for (a) ZnO, (b) MgO, and (c) GaSe, and demonstrate the enhancement in the dark-core size and reduction in the ring thickness of the harmonic vortices for increasing harmonic orders, in the transmission geometry. In (d) and (e), we present the results obtained from the reflection geometry, demonstrating the enhancement of the harmonic vortices beam size in Si and WSe$_2$, respectively. The OAM conservation law is demonstrated by counting the number of $2\pi$ phase shifts for harmonic orders $q=4$ and 5, as displayed in (f). In (a)-(e), $\beta_{x}$ denotes the far-field divergence along the $x$-axis, whereas, the horizontal, and vertical axes correspond to $\beta_{x}$, and $\beta_{y}$ in mrad unit, respectively for (f) (see Methods).}
\label{theory}
\end{figure}

For the description of the near-field harmonic vortices, which corresponds to harmonics generated exactly at the thin target's position, we consider that the emitted harmonics inherit the spatial field amplitude of the driving field. However, the harmonic amplitude scales with a factor, $p$, which is less than the harmonic order, $q$, in case of non-perturbative harmonics. For the perturbative harmonics, the scaling factor, $p$, coincides exactly with the harmonic order. The parameter, $p$, typically encodes information about the material, and its interplay with the TC of the driving field governs the far-field intensity distributions of different harmonics \cite{OAM_solids,Granados_2025}. Consequently, the near-field harmonic vortices can be regarded as scaled replicas of the fundamental field: $A_{q}^{near}(r',\phi') \propto |U(r')|^p \exp(-i q l_{0} \phi')$, where $A_{q}^{near}(r',\phi')$ represents the complex near-field amplitude of a harmonic vortex of order $q$, $U(r')$ is the spatial amplitude distribution of the driving LGV beam, and $\exp(-i q l_{0} \phi')$ is the helical phase of the $q$th-order harmonic. Moreover, Moreover, the transverse cartesian coordinates $x'$, $y'
$ are related to the polar coordinates by $r'=(x'^2+y'^2)^{1/2}$, and $\phi'=\arctan(y'/x')$, denoting the radial, and azimuthal coordinates of the vortex beam in the near-field, respectively. In order to extract $p$ values from the experimental measurements, we fit the measured changes in the harmonic vortex beam power as a function of the fundamental vortex beam power, both plotted in the logarithmic scale. The resulting values are presented in Table~\ref{TB1}.  

\begin{table}[ht!]
\centering
\caption{Scaling factor, $p$, for different harmonics in various materials.\label{TB1}}
\begin{tabular}{|c| c c c c c c|} 
\hline

Material & 3rd  & 4th  & 5th  & 6th  & 7th  & 9th  \\
\hline
ZnO      & 5.4 & 4.1   & 3.0  & 3.3  & 3.3  & –   \\
MgO      & 2.7 & –     & 5.3  & –    & 6.8  & –   \\
GaSe     & 3.5 & 3.4   & 4.3  & 4.4  & –    & –   \\
Si       & 0.9 & –     & 0.8  & –    & 0.9  & 0.8 \\
\hline
\end{tabular}
\end{table}

Table~\ref{TB1} reveals whether the generated harmonics are perturbative or non-perturbative in nature and shows how this character varies across materials. For instance, in ZnO, the fifth, sixth, and seventh harmonics clearly exhibit a non-perturbative characteristic since their scaling factor values are small when compared to their corresponding harmonic orders. However, for the third harmonic, the scaling factor, $p$, value is large as compared to the harmonic order making it difficult to determine the exact nature of the harmonic. Such a behavior of the third harmonic in ZnO for a driving wavelength of $1550$~nm was reported earlier in a theoretical work~\cite{Granados_2025}, and needs further investigation. Similarly, in GaSe, the fourth, fifth, and sixth harmonics are of non-perturbative nature, whereas, predicting the nature of the third harmonic is not straight-forward. In contrast, for the case of Si, all the generated harmonics are non-perturbative. Similarly, for MgO, the power scaling laws of the third and seventh harmonics clearly indicate a non-perturbative response, whereas, the fifth harmonic shows a scaling factor, $p$,value that is slightly higher than the harmonic order. For WSe$_2$, in contrast, it was not possible to measure the saturation curve to extract the exact $p$-value. We tested both perturbative, $p=3$, and non-perturbative, $p=2$, values and observed similar intensity distributions for the third harmonic.  

The knowledge of the $p$ values allows us to compute macroscopic far-field responses of solids, which is determined via the Fraunhofer diffraction theory; specifically, the far-field complex harmonic amplitude is calculated by applying the Fraunhofer diffraction integral to the complex near-field harmonic amplitude (see Methods). The resulting far-field line intensity profiles for different harmonics generated in the transmission geometry are presented in Figs.~\ref{theory} (a) to (c), whereas those obtained via the reflection geometry are shown in Figs.~\ref{theory} (d) to (e). Fundamentally, the material response, encoded in the scaling factor $p$, distinguishes the results obtained in the transmission geometry from their reflection counterparts. Most importantly, the theoretical model reproduces the primary experimental trends: the size of the dark-core of harmonic vortices increases with both the harmonic order $q$ and the driving field's TC $l_0$. Furthermore, the ring thickness of the harmonic vortices decreases with increasing harmonic order. These behavior, resulting from the nonlinear interaction between the vortex beam and the solid, is evident for harmonic orders $q=3,5$ and 7 in ZnO, although the model slightly overestimates the third-harmonic ring size. Moreover, for MgO and GaSe, the model successfully captures the general trends of increasing dark core-size and decreasing ring thickness with high-harmonic order, observed in the experimental data (see the Supplementary Material).

In the reflection geometry, the theoretical model also reproduces the experimentally observed enhancement in the harmonic beam size with the TC of the driving field, in agreement with the results obtained for Si and WSe$_2$. These effects originate directly from the conservation of OAM, implemented through the helical phase transformation, $\exp(i l_{0}\phi') \rightarrow \exp(i ql_{0}\phi')$, which naturally enforces the OAM conservation in the far-field (see Eq.~\ref{far}) and links the OAM up-scaling law to the spatial profile of the harmonic vortex beams. More importantly, the experimental verification of the OAM conservation via mode conversion, is accurately reproduced by the theoretical model, as shown in Fig.~\ref{theory} (f). These results show the far-field phase profiles for fourth and fifth harmonic vortices generated in ZnO. From Fig.~\ref{theory} (f), one can simply count the number of $2\pi$ phase shifts along the azimuthal coordinate to compute the TC of different harmonics (which is 4, and 5 for the fourth, and fifth harmonics, respectively). Across all investigated materials and geometries (transmission and reflection), the theoretical model reproduces the OAM conservation in solid-state HHG driven by vortex beams, following the well-established up-scaling law $l_q = ql_0$, consistent with gas-phase HHG~\cite{Corkun1}. 

\section*{Discussion}

The conservation of physical quantities and their tight connection to symmetries due to Noether's theorem \cite{Nother} plays also a central in light–matter interactions. As shown here, the OAM conservation law is preserved during the HHG process in different solid state materials. Additionally, the OAM of light provides a powerful additional degree of freedom for controlling and manipulating electronic, and structural dynamics in condensed-matter systems, where light–matter interactions are intrinsically richer than in atomic gases. Recent advances in helical dichroism have enabled OAM-dependent absorption, while the utilization of OAM to encode information in large Hilbert spaces—beyond the two degrees of freedom associated with electron spin—has become increasingly relevant for quantum technologies. Despite this progress, the role of light's OAM in non-perturbative strong-field interactions remains poorly understood, particularly in solid-state systems.

In this work, we advance the field by exploring the role of OAM in nonlinear, non-perturbative process. We generated femtosecond mid-infrared vortex beams and used to drive the HHG process in solids. Using a broadly applicable yet simple experimental approach, we show that the OAM of the emitted harmonics follows a strict linear up-scaling law, $l_q= q l_0$, identical to that observed in gas-phase HHG and consistent with a dipole-based description of the interaction. Most importantly, we demonstrate that the OAM conservation persists across a wide range of materials and HHG geometries, including systems with markedly different crystalline symmetries, band gaps, interaction mechanisms, and even strong spin–orbit coupling, as exemplified by monolayer WSe$_2$. The similar nonlinear response observed in both transmission and reflection geometries, together with the systematic increase in the dark-core size and reduction in the ring thickness of harmonic vortices with increasing harmonic order, indicates a common underlying mechanism governing the OAM transfer. We further show that propagation effects and dipole phase contributions do not modify the OAM up-scaling law. These observations allow us to treat both geometries within a unified framework, modeling the solid as an effectively-thin nonlinear medium, analogous to the thin-slab description commonly employed in atomic HHG. Within this picture, the microscopic response is primarily governed by the temporal driving field, while the spatial structure of the emitted harmonics is dictated by the vortex intensity profile of the driving beam. The calculated helical phase of the harmonics within this model accurately reproduces the experimentally observed OAM up-scaling, confirming that the conservation law is properly captured by this description.

Beyond establishing a fundamental conservation principle, our results render solid-state HHG as a versatile source of infrared-to-visible harmonic vortices with high TCs, directly generated in ambient air. To the best of our knowledge, the experimental generation of such high–TC harmonic beams from solids has not been previously reported. These sources may enable future studies of OAM-dependent light–matter interactions, including regimes beyond the conventional dipole selection rules, as well as applications requiring bright, structured ultrafast radiation.

Overall, our work shows that for the systems studied here, the OAM degree of freedom in solid-state HHG is governed by a robust conservation law consistent with a dipole-based description of strong-field light–matter interactions. By establishing this principle and introducing a simple, versatile source of high–TC harmonic vortices, our study lays the groundwork for engineering high–topological-charge vortex beams and extending ultrafast science toward structured attosecond radiation generated via solid-state HHG.

\section*{Methods}
\subsection*{Experimental details}

The size of the LGV beam at the focus was measured to be $w\approx 65~\mu$m, which is within the paraxial limit since the beam divergence is much less than 1 radian because the beam size is much larger as compared to the wavelength of the beam, $\lambda_{0}= 3.2~\mu $m. The beam size was measured by a scanning-slit laser beam profiler, also known as NanoScan. The fs MIR LGV beams were produced by imprinting the helical phase (therefore, OAM) onto the fs MIR Gaussian pulses\cite{Laser_System} using a spiral phase plate (SPP Vortex Photonics, Munich). The resulting vortex beams were then focused onto different target crystals using an off-axis parabolic mirror (OPM) with a focal length of $f_{l}=100$~mm. In order to detect harmonics in the ultraviolet spectral regime requires focusing the driving vortex beam at the exit face of the sample. After the HHG process, the driving field was filtered out by a borosilicate crown glass (BK7), and the generated harmonics were re-imaged by a lens with a focal length of $f=250$~mm, allowing us to record the harmonic vortices at the focal plane. The individual harmonics were filtered using band-pass filters (Thorlabs). To determine the TC of the harmonic vortices, we employed a cylindrical lens (CL)~\cite{Beijersbergen1993}. The CL operates via astigmatic mode conversion~\cite{Beijersbergen1993,Padgett2002}, transforming the input doughnut-shaped vortex mode into a tilted Hermite–Gaussian (HG) intensity distribution. From this HG intensity distribution, the TC, $l_q$, of the harmonic beam is retrieved by counting the number of intensity minima, $n$, along the diagonal of the intensity distribution. Specifically, the TC is directly given by the number of minima: $l_q = n$. Notably, the same experimental technique was implemented for measuring the fundamental vortex field's TC, $l_0$. 

Furthermore, the intensity distributions of the generated harmonics were captured with an imaging camera (Basler acA1440, spectral range 1200–300~nm). The spectral characterization of the harmonic vortices were performed with a spectrometer (AVANTES) with a spectral detection range from 1200 to 200~nm. 

\subsection*{Theoretical description of the fundamental LG vortex beams and modeling their nonlinear non-perturbative interaction with solids}

The simulation of the light-solid interaction with vortex beams in this work follows Ref.~\citenum{Granados_2025}.
The complex spatiotemporal field amplitude of the fundamental LGV beam (not to be confused with spatiotemporal vortex beams, where the spatial and temporal coordinates are intrinsically coupled, therefore, they possess transverse OAM) can be written as: 

\begin{eqnarray}
    U(r',\phi',z,t)=\frac{w_0}{w(z)}\Bigg(\frac{\sqrt2 r'}{w(z)}\Bigg)^{|l_{0}|} L_{P_{0}}^{|l_{0}|} \left(\frac{2r'^{2}}{w(z)^2}\right) e^{-\Big(\frac{r'}{w(z)}\Big)^{2}} e^{i l_{0} \phi'}  e^{\frac{i \kappa r'^{2}}{2R(z)}}e^{i\varphi_{G}(z)} E(t),
    \label{lgbeam}
\end{eqnarray}

 where $l_{0}$ is the longitudinal OAM of the LGV beam, $P_0$ is its radial index, $w_{0}$ is the Gaussian beam waist size, and $\varphi_G=-(2P_0+|l_{0}|+1)\arctan(z/z_R)$ is the Gouy phase of the LGV beam. Additionally, $R(z)=z[1+\left(z_R/z\right)^2]$, and $w(z)=w_{0}[1+\left(z/z_R\right)^2]^{1/2}$ represent the phase front radius, and the width of the beam at a finite propagation distance, $z$.  The Rayleigh range of the beam is given by $z_R=\kappa w_{0}^2/2$, with $\kappa=2\pi/\lambda_{0}$, and $\lambda_{0}$ representing the wave number, and wavelength of the beam, respectively. The temporal part of the electromagnetic field is given by: 

\begin{equation}
    \label{et}
    E(t)=E_0\sin^2\Bigg(\frac{\pi t}{n_c T}\Bigg)\sin(\omega_{0} t),
\end{equation}

where, $E_{0}$ is the peak electric field amplitude, $n_{c}$ denotes the total number of optical cycles in the laser field, $T=\lambda_{0}/c$ is the laser period, and $\omega_{0}=2\pi/T$ represents the central frequency of the driving laser field. We use the electrical field in Eq.~\ref{et} in the semiconductor Bloch equations (SBEs) to calculate the power scaling laws ($p$-value) for different harmonic orders \cite{Granados_2025}. Here, the $p$-values are extracted from the fitting of the harmonic beam power as a function of the driving laser beam power, plotted in the logarithmic scale. This implies that our model is now semi-empirical. Additionally, since we consider the dipole approximation to remain valid in our study due to a large disparity of spatial scales involved in the vortex light-solid interaction, the temporal dynamics can be separated from the beam's spatial structure. It implies that the nonlinearity of the process enters the theory via the $p$-values and the spatially structured part via the thin-slab model. This procedure avoids including the beam spatial structure into the solution of the SBEs, greatly simplifying the vortex light-solid interaction (macroscopic response) calculations. 
A direct consequence of this approach is that in the near-field, one can express the spatial amplitude distribution of an individual harmonic field as the fundamental field's spatial amplitude raised to the power of the scaling factor, $p$. Since the fundamental vortex beam exhibits a helical phase distribution, harmonic vortices are also generated with helical phases due to the coherent nature of the HHG process. In order to include the experimentally observed OAM conservation law, it is necessary to scale the helical phase of a given harmonic as the harmonic order times the helical phase of the fundamental field. Apart from the helical phase, the near-field harmonic's phase also includes scaled (with the harmonic order) Gouy and radial phases, and the intrinsic dipole phase. However, since we position the thin target exactly at the fundamental beam's focus position (i.e., at $z=0$), the contributions to the harmonic phase arising from the Gouy and radial phases are neglected. It is also important to highlight that if the target is placed at a position with $z\neq0$, the Gouy and radial phase terms can be included in the computation of both the near- and far-field complex harmonic amplitudes. The complex amplitude of a harmonic vortex in the near-field can mathematically be expressed as: 

\begin{eqnarray}
   A^{near}_{q}(r',\phi')\propto \big|U(r')\big|^p e^{i q \Phi(\phi')} \label{near},
\end{eqnarray}

Here, the term $\Phi(\phi')$ condenses all the field phases and $U(r')$ encapsulates the spatial amplitude terms of the fundamental LGV beam in Eq.~\ref{lgbeam}. Furthermore, in our computation, we consider a zero radial index ($P_{0}=0$), to ensure that there is a single bright ring in the transverse intensity distribution of the LGV beam. The propagation of the individual harmonic vortex beam from the near-field to the far-field can be obtained by using the Fraunhofer diffraction theory. This means that the near-field complex harmonic amplitude is propagated to the far-field by the Fraunhofer diffraction integral. After solving the angular integral analytically, the far-field harmonic vortex takes the following form: 

\begin{eqnarray}
    A_q^{far}(\beta, \phi)&\propto& e^{iql_{0}\phi}(-i)^{ql_{0}} \int_0^\infty r' dr' \big|U(r')\big|^pJ_{ql_{0}}\Bigg( \frac{2\pi\beta r'}{\lambda_q} \Bigg), \nonumber \\ \label{far}
\end{eqnarray}

where, ($\beta,\phi$) are the far-field coordinates representing the divergence, and azimuthal angle, respectively, $\lambda_{q}=\frac{\lambda_{0}}{q}$ is the wavelength of the $q$-th order harmonic, and $J_{ql_{0}}(...)$ is the Bessel function of the first kind of order $ql_{0}$. Depending on the type of vortex beam used for driving the HHG process, the solution of Eq.~\ref{far} can be found either analytically or numerically. In general, the integral results in a doughnut-shaped harmonic field with a helical phase that scales with the harmonic order, hence fulfilling the experimentally measured OAM up-scaling law. For the case of an LGV beam, the solution of Eq.~\ref{far} results in the following intensity distribution in the far-field \cite{Granados_2025}:

\begin{equation}    I(\beta_x,\beta_y)\propto\frac{I_0^p (1)^{ql_{0}}\left(\frac{2 \pi q \sqrt{\beta_x^2+\beta_y^2}}{    {\lambda_{0}}}\right)^{2ql_{0}}\Gamma \left[\frac{1}{2} (pl_{0}+ql_{0}+2)\right]^2}{2^{2ql_{0}+2}\left(\frac{p}{w_0^2}\right)^{pl_{0} +ql_{0}+2}\Gamma[ql_{0}+1]^2}
\times\left|{_1F_1} \Bigg[\frac{pl_{0}+ql_{0}+2}{2};ql_{0}+1;-\frac{\pi^2 q^2 w_0^2 \left(\beta_x^2+\beta_y^2\right)}{p\lambda_{0}^2} \Bigg]\right|^2, 
    \label{TIP}
\end{equation}

where, $I_{0}$ is the peak intensity of the fundamental LGV beam, $\beta=\sqrt{\beta_{x}^2+\beta_{y}^2}$, ${_1F_1}(...)$ is the confluent hypergeometric function, and $\Gamma(...)$ is the Euler-Gamma function. Basically, we utilize Eq.~\ref{TIP} to calculate the far-field harmonic intensity and Eq.~\ref{far} to calculate far-field phase distributions depicted in Fig.~\ref{theory}.

% \noindent LaTeX formats citations and references automatically using the bibliography records in your .bib file, which you can edit via the project menu. Use the cite command for an inline citation, e.g.  \cite{Hao:gidmaps:2014}.

% For data citations of datasets uploaded to e.g. \emph{figshare}, please use the \verb|howpublished| option in the bib entry to specify the platform and the link, as in the \verb|Hao:gidmaps:2014| example in the sample bibliography file.

\section*{Acknowledgments}
The ELI ALPS project (GINOP-2.3.6-15-2015-00001) is supported by the European Union and co-financed by the European Regional Development Fund. C.G. and B.K.D. acknowledge the support of the European laser Infrastructure ERIC through the program ``ELI Call for Users". C.R.G. acknowledge support from SECIHTI, México (CBF-2025-I-1804). We thank Andrew Forbes for the valuable comments on the manuscript. 

% Acknowledgements should be brief, and should not include thanks to anonymous referees and editors, or effusive comments. Grant or contribution numbers may be acknowledged.

\section*{Author contributions statement}
D.R., B.K.D. and R.S. contributed equally to this work. C.G., B.K.D., K.B., D.R. and E.C. conceptualized the idea. C.G. and W.G. prepared the materials for the experiments. C.G., D.R., R.S. and K.B. performed the experiments. C.G., C.R.G. and B.K.D. wrote the manuscript. C.G., W.G., S.F. and Q.Z. supervised the project. All authors reviewed and corrected the manuscript. 

\section*{Conflict of Interest Statement}
The authors declare no conflict of interest. 

\bibliography{sample}

% Must include all authors, identified by initials, for example:
% A.A. conceived the experiment(s),  A.A. and B.A. conducted the experiment(s), C.A. and D.A. analysed the results.  All authors reviewed the manuscript. 

% \section*{Additional information}

% To include, in this order: \textbf{Accession codes} (where applicable); \textbf{Competing interests} (mandatory statement). 

% The corresponding author is responsible for submitting a \href{http://www.nature.com/srep/policies/index.html#competing}{competing interests statement} on behalf of all authors of the paper. This statement must be included in the submitted article file.

% \begin{figure}[ht]
% \centering
% \includegraphics[width=\linewidth]{stream}
% \caption{Legend (350 words max). Example legend text.}
% \label{fig:stream}
% \end{figure}

% \begin{table}[ht]
% \centering
% \begin{tabular}{|l|l|l|}
% \hline
% Condition & n & p \\
% \hline
% A & 5 & 0.1 \\
% \hline
% B & 10 & 0.01 \\
% \hline
% \end{tabular}
% \caption{\label{tab:example}Legend (350 words max). Example legend text.}
% \end{table}

% Figures and tables can be referenced in LaTeX using the ref command, e.g. Figure \ref{fig:stream} and Table \ref{tab:example}.

\end{document}